\documentclass[11pt,letterpaper,amsmath,eqsecnum,
  amssymb,aps,prd,nofootinbib,
a4paper,superscriptaddress,altaffilsymbol]{revtex4}

\usepackage{amsmath}
\usepackage{amsfonts}
\usepackage{amssymb}
\usepackage{graphicx}
\usepackage{psfrag}
\newcommand{\bx}{{\bf x}}
\newcommand{\bk}{{\bf k}}

\newcommand{\Rs}{{R}}
\newcommand{\Hs}{{\mathcal{H}}}
\newcommand{\Rc}{{\mathcal{R}_{\rm c}}}

\newcommand{\be}{\begin{equation}}
\newcommand{\ee}{\end{equation}}
\newcommand{\ba}{\begin{eqnarray}}
\newcommand{\ea}{\end{eqnarray}}
\newcommand{\beq}{\begin{equation}}
\newcommand{\eeq}{\end{equation}}
\newcommand{\beqa}{\begin{eqnarray}}
\newcommand{\eeqa}{\end{eqnarray}}

\newcommand{\eref}[1]{Eq.~\eqref{#1}}

\newcommand*{\cref}[1]{Chapter~\ref{#1}}

\begin{document}

\title{Non-Gaussian initial conditions in $\Lambda$CDM: Newtonian,
  relativistic, and primordial contributions}
\author{Marco Bruni}
\email[]{marco.bruni@port.ac.uk}
\affiliation{Institute of Cosmology and Gravitation, University of
  Portsmouth, Dennis Sciama Building, Portsmouth PO1 3FX, United Kingdom.}

\author{Juan Carlos Hidalgo}
\email[]{hidalgo@fis.unam.mx}
\affiliation{Institute of Cosmology and Gravitation, University of
  Portsmouth, Dennis Sciama Building, Portsmouth PO1 3FX, United Kingdom.}
\affiliation{Instituto de Ciencias F\'{\i}sicas, Universidad Nacional
Aut\'onoma de M\'exico, Cuernavaca, Morelos, 62210, Mexico}

\author{Nikolai Meures}
\email[]{nmeures@gmail.com}
\affiliation{Institute of Cosmology and Gravitation, University of
  Portsmouth, Dennis Sciama Building, Portsmouth PO1 3FX, United Kingdom.}

\author{David Wands}
\email[]{david.wands@port.ac.uk}
\affiliation{Institute of Cosmology and Gravitation, University of
  Portsmouth, Dennis Sciama Building, Portsmouth PO1 3FX, United Kingdom.}

\date{\today}
\begin{abstract}

The goal of the present paper is to set initial conditions for
structure formation at non-linear order, consistent with general
relativity, while also allowing for primordial non-Gaussianity.  
We use the non-linear continuity and Raychaudhuri equations, which
together with the non-linear energy constraint determine the 
evolution of the matter density fluctuation in general relativity. 
We solve this equations at first and second order
in a perturbative expansion, recovering and extending previous results
derived in the matter-dominated limit and in the Newtonian regime.  
We present a second-order solution for the comoving density contrast
in a $\Lambda$CDM universe, identifying non-linear contributions coming from
the Newtonian growing mode, primordial non-Gaussianity and intrinsic
non-Gaussianity, due to the essential non-linearity of the
relativistic constraint equations. We discuss the application of these
results to initial conditions in N-body simulations, showing that
relativistic corrections mimic a non-zero non-linear parameter $f_{\rm
NL}$.     
\end{abstract}

\maketitle
%
\section{Introduction}\label{Ch6:sec:Intro}

The non-linear evolution of primordial fluctuations can be studied to learn
about the physics of the early universe. In this context, the
bispectrum of the cosmic microwave background radiation (CMB) and its
relation to primordial non-Gaussianity has become an important tool to
study the conditions in the inflationary universe
\citep{Bar04,Wands:2010af,Koyama:2010xj,Byrnes:2010em}. More recently,
galaxy surveys have been used as a means to constrain primordial
non-Gaussianity in large-scale structure (LSS)
\citep{Dalal:2007cu,Matarrese:2008nc,Desjacques:2010jw,Giannantonio:2013uqa}.   

Extracting information related to the
primordial non-Gaussianity from the evolved matter fluctuations is,
however, not trivial. The matter fields evolve non-linearly under the
action of gravity and thus, even if the primordial power spectrum is
completely Gaussian, non-linear processes would still introduce
non-Gaussian correlations \citep{MatMolBru98,Bernardeau:2001qr,Bar05,Bar10}.   
It is therefore essential to understand which kind
of non-Gaussianities are induced in the matter distribution by
gravity and other physical processes in order to be able to
reconstruct non-Gaussianities in the primordial field.  

Given the complexity of the full non-linear equations in general
relativity (GR), Newtonian equations are typically used both in
analytic treatments of the non-linear growth of structure
\citep{Pee80,Bernardeau:2001qr} and in numerical (e.g., N-body)
simulations, even when initial conditions are set on scales far larger
than the causal horizon in the matter-dominated universe. This leads
to the {open} question of how to interpret Newtonian results from a
relativistic perspective, see
e.g.\ \cite{Chisari:2011iq,Green:2011wc}, 
{c.f.\ also \cite{flender:2013a,Haugg:2012ng}}.   
In this paper we consider the relation between Newtonian dynamics of
matter perturbations in a $\Lambda$CDM universe and the full general
relativistic dynamics at non-linear level. It is well known that
linear matter perturbations in general relativity obey the same
evolution equations as in Newtonian theory, and it has previously been
argued that Newtonian equations accurately reproduce GR evolution up
to third order \citep{Noh:2004bc}, however we show that there are
important non-linear constraint equations which lead to corrections to
the initial conditions at non-linear orders in the perturbative
expansion \citep{Bar05,Bar10}.   

We shall study the evolution of matter density perturbations for an
irrotational flow in $\Lambda$CDM, presenting the solution to the GR
equations for the density perturbations up to second order in the
synchronous-comoving gauge. This solution is consistent with previously known
separable solutions in the particular case of a matter-dominated
universe ($\Omega_m=1$)
\citep{tomita:67,MatMolBru98,Bar05,Hwang:2012bi}. {For $\Lambda$CDM, our
solution generalises those by \cite{Bar10} and \cite{Tomita:2005et}:
\citet{Bar10} solution is restricted to planar-symmetry (leading to
the pancake of the Zel'dovich approximation) in its Newtonian part;
 \citet{Tomita:2005et} does not explicitly relate the second
order solution to primordial non-Gaussianity}.   

The derivation of the second-order solutions in GR is necessarily
rather lengthy and technical in nature so we provide here an overview
of our main results. 

Firstly, in section \ref{Ch6:sec:synch}, we present the non-linear
evolution equations for inhomogeneous matter in a synchronous-comoving
gauge. By choosing observers comoving with the fluid four-velocity we
are following the spirit of the covariant fluid flow approach to
cosmology \citep{Ell71,maartens:2012}, which makes the relativistic
evolution almost identical to Lagrangian perturbations in Newtonian
theory. Indeed, our starting point is the standard system of coupled
evolution equations for the non-linear density perturbation, $\delta$,
and the inhomogeneous expansion of the matter flow, $\vartheta$.  

In section \ref{Ch6:sec:pert} we perform a perturbative expansion,
expressing the expansion, shear and 3-Ricci curvature in terms of
spatial metric perturbations. We derive the matter density
perturbations in synchronous-comoving gauge up to second perturbative
order. 

\begin{itemize}
\item
We first present the well-known first-order solutions to the
linearised evolution equation, identifying a first-integral of the
coupled evolution equations using the linearised energy
constraint. This shows how the 3-Ricci curvature scalar  $R^{(1)}$, 
which is constant, drives the growing mode of the density contrast
\beq
\delta^{(1)} \propto R^{(1)}({\bf x}) D_+(\tau) \,,
\eeq
where the linearised momentum constraint ensures that $R^{(1)}$ is a constant.
\item
We then repeat the same approach at second-order, using the energy
constraint and the second-order 3-Ricci scalar,
$R^{(2)}$. The 3-Ricci scalar is no longer constant at second order,
but its time dependence is given by the momentum constraint. We split
the 3-Ricci scalar into a constant part, $R^{(2)}_h$, and a
time-dependent part, $R^{(2)}_p$ such that
$R^{(2)\prime}_p=R^{(2)\prime}$. We thus split the second-order
solution into a homogeneous part, obeying the same first-integral as
at first order, with a growing mode solution 
\beq
\delta^{(2)}_h \propto R^{(2)}_h({\bf x}) D_+(\tau)
\,,
\eeq
and a particular solution, including time-dependent driving terms, quadratic in
first-order variables, 
\beq
\delta^{(2)}_p \propto \left( R^{(1)}({\bf x}) \right)^2
D_{2+}(\tau,\Sigma({\bf x})) \,.
\eeq
This solution is not in general separable since the growing mode is a
function of the shape parameter, $\Sigma$ defined in \eref{sigma:def},
which is in general inhomogeneous. For planar symmetry ($\Sigma=1$) or in the
matter-dominated limit ($\Omega_m=1$) the solution is separable and we
find $D_{2+}\propto(D_+)^2$. This particular growing mode then
dominates over the homogeneous solution at late times. 
\end{itemize}

In section~\ref{Ch6:sec:NGR} we discuss the relation to Newtonian results: 
\begin{itemize}
\item
The first-order growing mode for the density perturbation, $\propto
D_+$, whose amplitude is proportional to the the comoving curvature
perturbation $\Rc$ in general relativity, is known to coincide with
the first-order Newtonian solution, whose amplitude is proportional to
the initial Newtonian potential $\Phi_{IN}=(3/5)\Rc$.  
\item
The particular second-order growing mode solution, $\propto D_{2+}$,
whose amplitude is proportional to $\Rc^2$, reproduces the Newtonian
growing mode at second order in Lagrangian perturbation theory. We
show in an appendix that a spatial gauge transformation to Eulerian
coordinates reproduces the standard Eulerian density perturbation in
the matter-dominated limit. This leads to a growing non-Gaussianity,
but a bispectrum that vanishes in the squeezed limit. 
\item
The amplitude of the second-order homogeneous solution, $\propto D_+$,
is set by initial conditions. In Newtonian theory the Poisson equation
provides a linear relation between the density and Newtonian potential
at all times and at all orders in a perturbative expansion. Therefore
the initial second-order density perturbation is set by the initial
second-order potential, determined, for example, by the non-linearity
parameter $f_{\rm NL}$. The second-order homogeneous solution in
Newtonian theory is therefore due solely to primordial non-Gaussianity
in the potential. In GR we have non-linear constraint equations,
therefore the second-order homogeneous solution is non-zero in
general. Even in the absence of primordial non-Gaussianity, in the
squeezed limit we find an effective value of $f_{\rm NL}=-5/3$. 
\item
Newtonian simulations of structure formation can reproduce the
``true'' GR results up to second order, but care needs to be taken in
setting initial conditions and in identifying Newtonian and
relativistic variables.  
Respecting the non-linear constraint
equations of general relativity, including both primordial
non-Gaussianity and intrinsic non-Gaussianity, we can express
the initial ($\Omega_{mIN}=1$) density perturbation as: 
\beq  
 \label{gr:poisson1+2}
\delta_{IN} = \frac{2}{3\Hs^2_{IN}} \nabla^2 \Phi_{IN} -\frac{4}{3\Hs^2_{IN}} 
\left[  \left( f_{\rm NL} - \frac53 \right) \Phi_{IN}\nabla^2\Phi_{IN} +
\left( f_{\rm NL} + \frac{5}{12} \right) \partial^j\Phi_{IN} \partial_j\Phi_{IN}
 \right] \,.
\eeq

\end{itemize}

We conclude in Section~\ref{Ch6:sec:sum}.
%
\section{The non-linear analysis in synchronous-comoving
  gauge}\label{Ch6:sec:synch}  
%
We will calculate the evolution of the inhomogeneous matter density in
the synchronous-comoving gauge. In this section we will present the
exact, non-linear equations. Previous derivations have been discussed
in the context of a pure dust cosmology \citep{MatTer96,MatMolBru98,Bar05}
and $\Lambda$CDM \citep{Bar10}. 

From a general metric, written using conformal time $\tau$,
\beq
ds^2  = a^2(\tau) \left[- (1 + 2\phi) d\tau^2 
+ 2{\omega}_i d\tau dx^i  + 
 \gamma_{ij} dx^i  dx^j\right], 
\label{Ch6:def:metric}
\eeq

\noindent the synchronous gauge \citep{1975Landau} is defined by setting 
$\phi={\omega}_i=0$.
This implies that for every observer at a fixed spatial coordinate point of
the perturbed spacetime the proper time is the same as the cosmic time
in the FLRW background with scale factor $a(\tau)$. The synchronous
line element is therefore written in the form   
\beq
\label{Ch6:eq:linesync}
ds^2=a^2(\tau)\left[-d\tau^2+\gamma_{ij}(\bx,\tau)dx^idx^j\right],
\eeq 

\noindent where $\gamma_{ij}(\bx,\tau)$ is
the 3-metric.

We consider irrotational dust flow and choose observers comoving with
the fluid.   
This choice implies that the fluid 4-velocity can be made orthogonal
to the constant time spatial hypersurfaces with metric
$\gamma_{ij}$, that is, the four velocity in this gauge is
$u_{\mu}=[-a,0,0,0]$. The choice of such comoving observers is the
basis of the covariant fluid approach to perturbation theory
\citep{EllBru89,BruDunEll92}; here we follow the spirit of this
approach and take advantage of the simplifications implicit in the
choice of a set of coordinates that are simultaneously synchronous and
comoving; each fluid element has coordinates assigned by its initial
position \citep[c.f.][]{MatTer96}. As we shall see, this approach makes
the relativistic description almost identical to that of the
Lagrangian perturbation theory in the Newtonian context. Actually,
given that we also have a universal time, we may as well call our
gauge a Lagrangian gauge \citep{Villa:2011vt}.  

The starting point of our fluid-flow description is the deformation
tensor, defined as  
\beq
\vartheta^\mu_\nu \equiv {au^\mu}_{;\nu}-\Hs \delta^\mu_\nu\,, 
 \label{def:vartheta}
\eeq
\noindent where we have subtracted from the derivative of the
four-velocity the isotropic background expansion,  
given by the conformal Hubble scalar  $\Hs = a' / a$, where
a prime denotes the conformal time derivative.  
This deformation tensor plays a key role in our approach as it is
ubiquitous in the equations relevant to our study. Its trace 
$ \vartheta={\vartheta ^\alpha}_\alpha$,  represents the inhomogeneous
part of the volume expansion. The traceless part is the matter shear
tensor $\sigma^\alpha_\beta$, which represents the deviation from isotropy.    

Since we have chosen $u^\mu$ such that it
coincides with the normal to the constant time hypersurfaces,
the deformation tensor (\ref{def:vartheta}) is purely spatial and
coincides with the extrinsic curvature of the constant time slices in
the conformal space-time,
\beq 
\label{Ch6:def:curv}
\vartheta^i_j=-K^i_j\,,
\eeq 
where $K^i_j$ can be written as \citep{Wal84}   
\beq 
K ^i_j \equiv - \frac{1}{2}\gamma^{ik}\gamma'_{kj}\,.  
\eeq 

The continuity equation for dust follows from the energy
conservation equation $u_\alpha{T^{\alpha\beta}}_{;\beta}=0$, that is  
\beq
\label{Ch6:eq:conserv}
\frac{\rho'}{\rho}=-\frac{1}{2}\gamma^{ij}\gamma'_{ij}-3\Hs
=-\vartheta-3\Hs\,,
\eeq  

\noindent where $\rho$ is the total matter density. Formally we can solve \eref{Ch6:eq:conserv} in terms of the determinant,
$\gamma={\mathrm{det}}\left(\gamma^{ij}\right)$, to find   
\beq
 \rho=\frac{A(\bx)}{a^3\sqrt{\gamma}}\,.
\label{rho:gamma}
\eeq 

In the perturbative analysis we define the usual density contrast, $\delta$, by  
\beq 
\rho(\bx,\tau) = \bar{\rho}(\tau) + \delta\rho(\bx,\tau) =
\bar{\rho}(\tau) (1 + \delta(\bx,\tau))\,,
\eeq 

\noindent where $\bar\rho$ is the background density. Then the
continuity equation for the density contrast is 
\beq
\delta' + (1 + \delta)\vartheta = 0\,.
\label{nonlinear:continuity}
\eeq
for which the solution is  
\beq 
\label{Ch6:eq:deltaex}
\delta(\bx,\tau)=\frac{\delta_0(\bx)+1}{\sqrt{\gamma(\bx,\tau) /
    \gamma_0(\bx)}}-1\,.
\eeq 
This is the exact density fluctuation as a function of the metric,
without any approximation.  
In practise, however, one needs to solve Einstein's equations in order
to determine $\delta$ through $\gamma$. 

The evolution of $\vartheta$ is given by the Raychaudhuri equation
\beq 
\label{Ch6:eq:Ray} 
\vartheta'+\Hs \vartheta
+\vartheta^i_j\vartheta^j_i+
4\pi Ga^2\bar\rho\delta=0\,.
\eeq 
%
We note that the two
non-linear equations {(\ref{nonlinear:continuity})}
and~(\ref{Ch6:eq:Ray}) are formally identical to the Newtonian ones in 
the Lagrangian formalism \citep{Ell71,Pee80} (see discussion in
Sec.~\ref{Ch6:subsec:equiv}).   

In the relativistic case, Eq.\ (\ref{Ch6:eq:Ray})
can be obtained directly from a simple geometrical identity between
the deformation tensor $\vartheta^i_j$, and the 4-dimensional
curvature \citep{Ell71,Wal84,maartens:2012},  
after contraction and substitution from 
the 00-component of the Einstein
field equations (the energy constraint) 
\beq 
\label{Ch6:eq:evol}
\vartheta^2-\vartheta^i_j\vartheta^j_i+4\Hs\vartheta+\Rs =
16\pi Ga^2\bar\rho\delta\,,
\eeq  

\noindent where $\Rs$ is the trace of $\Rs^i_j$, the Ricci-curvature
of the 3-metric 
$\gamma^i_j$. Here we find it more useful to start from the
$ij$-component of the Einstein equations, which can be written as
\citep{Wal84,MisThoWhe73,niko:thesis} 
\beq 
\label{Ch6:eq:evol2}
{\vartheta^i_j}'+2\Hs\vartheta^i_j+\vartheta\vartheta^i_j +
\frac{1}{4}\left(\vartheta^k_l\vartheta^l_k-\vartheta^2\right)\delta^i_j
+ \Rs^i_j-\frac{1}{4}\Rs\delta^i_j=0\,,
\eeq

\noindent i.e., an evolution equation for $\vartheta^i_j$, which we will use later.
Combining the trace of Eq.\ (\ref{Ch6:eq:evol2}),  
\beq
\label{3Rid:trace}
\vartheta' +  2\Hs \vartheta + \frac{1}{4} \left[\vartheta^2 + 
  3 \vartheta^i_j \vartheta^j_i + \Rs \right]= 0,
\eeq
with the energy constraint \eref{Ch6:eq:evol}
\noindent to eliminate $\Rs$, one obtains the dynamical
equation~(\ref{Ch6:eq:Ray}).

%
The evolution of the deformation tensor (\ref{Ch6:eq:evol2}) is again
similar to its Newtonian counterpart, except that instead of second
derivatives of the Newtonian gravitational potential we now have the
Ricci tensor for the 3-metric. 

 
Finally, we note that the 0j-component of the Einstein field
equations yields the momentum constraint   
\beq 
\label{Ch6:eq:momcon} 
\vartheta^i_{j|i}=\vartheta_{,j}\,,
\eeq
\noindent where a stroke denotes a covariant derivative in the
3-space with metric $\gamma_{ij}$.

%
%

%
\section{The perturbative treatment}\label{Ch6:sec:pert}
%
The above treatment has been exact in the sense that we did not make
any assumption about the 3-metric $\gamma_{ij}$. 
Now we consider the case where the line element in
\eref{Ch6:eq:linesync} is close to the spatially flat FLRW and hence
we decompose $\gamma_{ij}$ up to second perturbative order 
\beqa
\gamma_{ij}&=&\delta_{ij}+\gamma^{(1)}_{ij}+\frac12
\gamma^{(2)}_{ij}+...
\nonumber \\ 
&=&
\left(1-2\psi^{(1)}-\psi^{(2)}\right)\delta_{ij} +
\chi^{(1)}_{ij}+\frac{1}{2}\chi^{(2)}_{ij}+...\,, \label{Ch6:eq:3met}
\eeqa 

\noindent where the superscript denotes the order of the perturbations
and 
\beq 
\chi_{ij}=\left(\partial_i\partial_j -\frac{1}{3}
\delta_{ij}\nabla^2\right) \chi\,,  
\label{chi:def}
\eeq

\noindent where $\nabla^2 = \delta_{ij}\partial^i \partial^j$. Similarly 
\beq
\delta = \delta^{(1)} + \frac12 \delta^{(2)} + \ldots,
\eeq

\noindent and equivalently for $\vartheta$ and $\Rs$.
Note that we have only included scalar quantities and that there are
only two scalar degrees of freedom at each order. Vector and tensor degrees of freedom
are linearly independent at first order and can consistently be set to zero, in which case they do not affect the second order scalar perturbations. Even if they do exist at first order, they don't affect first-order density perturbations and at second order their effect is subdominant \citep{MatMolBru98,MalWan09}.   



Using the equations of Sec.\ \ref{Ch6:sec:synch}, we
can construct differential equations governing the growth of
fluctuations at each perturbative order. 
%
\subsection{The background dynamics}
  In a spatially flat $\Lambda$CDM cosmology the energy constraint gives
  the Friedmann equation; in terms of $\Hs = a'/a = a H$ this is
\beq
3 \Hs^2  =  8\pi G a^2 \bar\rho + a^2\Lambda.
\label{friedmann:eq}
\eeq

\noindent The Raychaudhuri equation  is 
\beq
\label{Ray:bg}
3 \Hs' + 4\pi G a^2\bar\rho  - a^2 \Lambda= 0,
\eeq
\noindent and the homogeneous part of the continuity equation
\eref{Ch6:eq:conserv} is simply
\beq
\label{barrho:eq}
\frac{\bar\rho'}{\bar\rho} = - 3 \Hs.
\eeq

 These equations determine the dimensionless density 
parameter $\Omega_m \equiv 8\pi G a^2\bar{\rho} / 3 \Hs^2$ in terms of the
scale factor, 
\beq
\Omega_{m}(\tau) = \Omega_{m0}\left[\Omega_{m0} + a(\tau)^3(1 -
  \Omega_{m0}) \right]^{-1}\,,
\label{Omega:def}
\eeq
 
\noindent with the current value $\Omega_{m0} = \Omega_{m}(\tau_0)
\simeq 0.315 \pm 0.017$ \citep{planck:param}.  

Note finally that combining Eqs.~\eqref{friedmann:eq} and
\eqref{Ray:bg} one obtains a homogeneous equation for the Hubble
expansion $H$; this can be cast as 
\beq
\label{hubble:equation}
{H}' + \frac32 \Hs \Omega_m H  = 0\,, 
\eeq

\noindent and in this form it  will be useful in Sec.~\ref{Ch6:sec:first}. 
%
\subsection{Kinematical, curvature and metric variables at first order}
\label{sec:kincurv}
%
Expanding at first order the quantities introduced in
Sec.\ \ref{Ch6:sec:synch} we can relate them to the scalar metric
potentials $\psi^{(1)}$ and $\chi^{(1)}$, Eqs.~(\ref{Ch6:eq:3met})
and~(\ref{chi:def}) \citep{BruDunEll92,MatMolBru98}.   
The deformation tensor \eref{Ch6:def:curv} is given by
\beq 
\label{Ch6:eq:extr11}
{\vartheta^{(1)i}}_j=-{\psi^{(1)}}'\delta^i_j+\frac{1}{2}
\left({\chi^{(1)}}_j^i\right)'\,.
\eeq 

\noindent The trace and traceless parts of $\vartheta^{(1)i}{}_j$
are, respectively, the
inhomogeneous expansion scalar and the matter shear:  
\begin{align}
\label{theta:trace}
 \vartheta^{(1)}=& -3{\psi^{(1)}}'\,,\\
\label{theta:traceless}
{\sigma^{(1)}}^i_{j}{} =& \frac{1}{2} \left({\chi^{(1)}}_j^i\right)'\,.
\end{align}

\noindent Additionally, expanding the 3-Ricci scalar  at
first order one gets
\beq
\label{R1:expansion}
\Rs^{(1)} = 4 \nabla^2 \left[\psi^{(1)} +
  \frac{1}{6}\nabla^2\chi^{(1)}\right] \,. 
\eeq

Note that the matter shear $\sigma^{(1)}{}^i_j$ and the 3-Ricci
curvature of the comoving orthogonal hypersurfaces 
$\Rs^{(1)}{}^i_j$ are 
tensors that vanish in the background (the latter only in a flat background) 
and as such, they are gauge-invariant quantities, represented in our
synchronous-comoving gauge  by the RHS of Eqs.~\eqref{theta:traceless}
and~\eqref{R1:expansion} \citep{BruDunEll92,Bru96}.  
In this gauge, the above relations show that $\vartheta$ and
$\sigma^i_j$ coincide with the expansion and the matter shear of the
normal to the time-slicing\footnote{Note that the shear and the
  3-Ricci curvature of an arbitrary slicing are not gauge-invariant
  \citep{KodSas84,MalWan09}.}.  

An important quantity in the relativistic perturbation theory
of the early universe is  the comoving curvature perturbation, $\Rc$, the conformally flat part of
the metric perturbation on comoving hypersurfaces
\citep{Lyth:1984gv}. In terms of our metric variables, Eqs.~(\ref{Ch6:eq:3met}) and~(\ref{chi:def}), this is    
\beq 
\Rc = \psi^{(1)} + \frac16 \nabla^2 \chi^{(1)}\,,
\label{R1c:expansion}
\eeq  

\noindent so that, 
from \eref{R1:expansion},
\beq
\Rs^{(1)} = 4 \nabla^2 \Rc. 
\label{R1:Rc1}
\eeq

We can also use the gauge-invariant potentials $\Phi$ and $\Psi$
\citep{Bar80,MalWan09}, which coincide with $\phi^{(1)}$ and
$\psi^{(1)}$ in the Poisson gauge. In our gauge,  
\begin{align}
\Phi =&\, - \frac12 \left(\chi^{(1)}{}'' + \Hs \chi^{(1)}{}'\right),
\notag\\
\Psi =&\, \psi^{(1)} + \frac16 \nabla^2 \chi^{(1)} + \frac12 \Hs
\chi^{(1)}{}'\,. 
\label{phi:def}
\end{align}

\noindent  From the vanishing of the anisotropic stresses in 
Einstein's equations it follows that $\Phi = \Psi$ \citep{Bar80}. 

Finally, we remark that the gauge-invariant potential $\Phi$ and the
first-order density perturbation in our comoving gauge are related
through the Poisson equation \citep{Bar80,BruDunEll92,Wands:2009ex},  
\beq 
\label{Ch6:eq:Poi}
\nabla^2\Phi=\frac{3}{2}\Hs^2\Omega_m\delta^{(1)}\,.
\eeq

\noindent Another important quantity in studies of the early universe is the gauge-invariant curvature perturbation on uniform-density hypersurfaces \citep{MalWan09}. This is given by
\beq
\label{zeta:def}
\zeta^{(1)} \equiv 
- (\psi^{(1)} + \frac{1}{6}\nabla^2\chi^{(1)}) -  \frac{\Hs}{\rho'} \delta\rho^{(1)}
= - \Rc + \frac{1}{3} \delta^{(1)} \,. 
\eeq

\noindent 
We note from \eref{Ch6:eq:Poi} that $\delta^{(1)}$ is suppressed on large scales, well outside the horizon,
and therefore, at early times
\beq
\label{zeta:R}
\zeta^{(1)} \simeq -\Rc.
\eeq

\noindent In the rest of this section, we use $\Rc$ to express
our initial conditions in terms of this gauge-invariant
quantity\footnote{$\delta^{(1)}$ and 
  $\vartheta^{(1)}$ are not themselves gauge-invariant quantities, but
  they represent gauge-invariant variables when evaluated in our
  gauge \citep{MalWan09}.}.     

Since our goal is to set initial conditions for structure formation at
some early initial time $\tau_{IN}$ in the matter-dominated era,
various quantities evaluated at this time will be indicated with the
sub-index $IN$. 
%
\subsection{First order solutions}\label{Ch6:sec:first}
%
We start by writing the first-order part of the continuity equation
\eqref{Ch6:eq:conserv} as    
\beq 
\label{Ch6:eq:consfirst}
{\delta^{(1)}}' + \vartheta^{(1)} = 0\,.
\eeq 

\noindent The first-order expansion of the Raychaudhuri equation
\eqref{Ch6:eq:Ray} takes the form  
\beq 
\label{Ch6:eq:Rayfirst}
{\vartheta^{(1)}}'+\Hs \vartheta^{(1)} +\frac32 \Hs^2
\Omega_m\delta^{(1)} = 0\,.
\eeq

\noindent We thus obtain two equations for $\delta^{(1)}$ and
$\vartheta^{(1)}$ which are decoupled from other perturbations at
first order. 
Therefore the solutions of the above equations solve the problem of
the first-order matter density evolution. Furthermore, since there are
only two scalar degrees of freedom, all other perturbations can be
expressed in terms of these solutions.  

Combining equations \eqref{Ch6:eq:consfirst} and
\eqref{Ch6:eq:Rayfirst} we obtain the evolution equation for the
density contrast
\beq
\label{Ch6:eq:deltadd2}
      {\delta^{(1)}}''+\Hs{\delta^{(1)}}' -
      \frac{3}{2}\Hs^2\Omega_m\delta^{(1)} = 0 \,.
\eeq 

\noindent  This is the same as the Newtonian evolution equation for the
matter density fluctuation \citep{Pee80}. 
 

In the relativistic formalism, we can use the energy constraint
\eqref{Ch6:eq:evol} to make a direct link with early universe 
fluctuations in terms of the 3-Ricci scalar $\Rs$. At first order, the
energy constraint 
\eqref{Ch6:eq:evol} yields an algebraic relation between the three
variables $ \vartheta^{(1)},\, \Rs^{(1)}$ and $\delta^{(1)}$: 
\beq
\label{energy:const1}
4 \Hs \vartheta^{(1)}  - 6 \Hs^2 \Omega_m \delta^{(1)} + \Rs^{(1)}= 0 \,.
\eeq

\noindent Using
\eref{Ch6:eq:consfirst} to eliminate $\vartheta^{(1)}$ we find 
\beq
4 \Hs {\delta^{(1)}}' + 6 \Hs^2 \Omega_{m}\delta^{(1)}  
- \Rs^{(1)} = 0.
\label{first:intdelta}
\eeq

On the other hand, taking the time derivative of \eref{energy:const1},
and eliminating ${\delta^{(1)}}'$ and ${\vartheta^{(1)}}'$ using
Eqs.~\eqref{Ch6:eq:consfirst} and \eqref{Ch6:eq:Rayfirst}, gives 
\beq
{\Rs^{(1)}}' = 0\,.
\label{R:const}
\eeq

Hence,  \eref{first:intdelta}  is a first integral of the evolution
equation \eref{Ch6:eq:deltadd2} where the 3-Ricci scalar
$\Rs^{(1)}$ is a constant to be determined by initial conditions.
Note that the momentum constraint \eref{Ch6:eq:momcon} gives, at
first order,
\beq 
\label{Ch3:eq:needlabel} 
\partial _j\left( 6\psi^{(1)}  + \nabla^2\chi^{(1)}\right)'=0\,,
\eeq 

\noindent i.e., $ 6\, \partial_j {\Rc}'= 0$.
Therefore, since \eref{R:const} implies that $\Rc = \mathrm{const.}$,
the momentum constraint is identically satisfied, which shows the
consistency of our procedure. In addition, this constraint implies that 
$\psi^{(1)}{}' = - \frac{1}{6}\nabla^2 \chi^{(1)}{}'$,
which allows us to write
\beq
\label{theta1:chi1}
\vartheta^{(1)}{}^i_{~j} = \frac12  \partial^i\partial_j\chi^{(1)}{}'\,.
\eeq

\noindent We will use this last result to simplify our second-order
calculations. 

It is standard practice to write the general solution of \eref{Ch6:eq:deltadd2}
as a linear combination of a growing mode and a decaying mode:  
\beq
\delta^{(1)}(\tau,\bx) = \delta_+^{(1)}(\bx) D_+(\tau) +
\delta_-^{(1)}(\bx) D_-(\tau)\,.
\eeq 

We can relate these solutions with $\Rs^{(1)}$ through the first 
integral \eref{first:intdelta}. The decaying mode $D_- $
is the solution to the homogeneous part of \eref{first:intdelta}
\beq
D_-' + \frac32 \Hs \Omega_{m}D_- = 0,
\eeq

\noindent i.e., the decaying mode $D_{-}$ is associated with isocurvature
perturbations. Comparing with \eref{hubble:equation} immediately gives
the solution $D_- = D_{-IN}H / H_{IN}$. 
The growing mode instead corresponds to the particular solution of the first
integral \eref{first:intdelta}, with $ \Rs^{(1)} \neq 0$ and a specific
initial condition $\delta_{IN}$ related to $\Rs^{(1)}$ itself.
This explicitly shows how the curvature perturbation drives the
formation of structure. Hereafter, we discard the decaying mode and in
what follows we write     
\beq
\label{delta:decom}
\delta^{(1)}(\tau, \bx)=C_1(\bx) D_+(\tau)\,, 
\eeq

\noindent where $C_1(\bx) = \delta^{(1)}_+(\tau_0,\bx)$,
i.e., $D_+(\tau_0) = 1$, and the growth factor $D_+(\tau)$
corresponds to the particular solution of \eref{first:intdelta} written as 
\beq 
\label{dplus:integral}
C_1(\bx)\left[\Hs D'_+ + \frac32 \Hs^2 \Omega_{m} D_+\right] - 
\frac14 \Rs^{(1)} =0\,.
\eeq 

Since $\Rs^{(1)}$ is a constant, the square bracket is also a
constant, which allow us to express $C_1$ in terms of
$\Rs^{(1)}$ or $\Rc$:
\beq
\label{delta0:R}
C_1  = \frac23 \frac{\nabla^2 \Rc}{\Hs^2_{IN}\Omega_{m IN}} \left[1 +
    \frac23 \frac{f_1(\Omega_{m IN})}{\Omega_{m IN}}\right]^{-1}\frac{1}{D_{+ IN}} ,
\eeq

\noindent where one can define \citep{Pee80}
\beq 
\label{Ch6:eq:fm} 
{f_1 \equiv \frac{D_+'}{\Hs D_+} = \frac{d \log D_{+}}{d \log a} =
-\frac{3}{2}\Omega_m + \frac{\Omega_m a}{\delta^{(1)}}
\frac{\Rs^{(1)}}{4 \Hs_0^2 \Omega_{m 0}}  \,,}
\eeq

\noindent {and the last equality is obtained using \eref{first:intdelta}.
Note that, given that $\Omega_m$ is a monotonic function of $a$,
\eref{Omega:def} can be inverted, so that $f_1 = f_1(\Omega_m)$.} In a
matter-dominated universe with $\Omega_{m}=1$ we have $D_+\propto a
\propto \tau^2$ and hence $f_1(1) = 1$. More generally, from
\eref{dplus:integral} we have 
\beq
\label{expressD+}
\left[ f_1(\Omega_m) + \frac32 \Omega_{m} \right] \Hs^2 D_+ = {\rm const} \,,
\eeq
and hence
\beq
\label{Dplus:soln}
\frac{D_+}{D_{+IN}} = \left[ \frac{5}{2f_1(\Omega_m)+3\Omega_m}
  \right] \frac{\Hs_{IN}^2}{\Hs^2} \,. 
\eeq
In $\Lambda$CDM, while the universe is still matter dominated,
$\Omega_m\simeq1$; { assuming $ f_1 = \Omega_m^{q}$ and ${1 - \Omega_m  \ll
1}$ we find, from the time derivative of the first integral \eref{expressD+}}
\beq 
\label{f1:Omegam}
f_1(\Omega_m) \simeq \Omega_m^{6/11},
\eeq

\noindent {i.e. {$q = 0.\overline{54}$}, in agreement with
\cite{Wang:1998gt} and \cite{Linder:2007hg}. Different approximations lead to
slightly different values of $q$
\citep{Bernardeau:2001qr,Carroll:1991mt,Lahav:1991wc}, but the difference
is negligible at the early times considered here.}  Henceforth we will
set initial conditions at early times in the matter-dominated era,
$\Omega_{m IN}=1$, such that \eref{delta0:R} simplifies to  
\beq
C_1 = \frac25 \frac{\nabla^2 \Rc}{\Hs^2_{IN}} \frac{1}{D_{+ IN}} .
\label{C1early}
\eeq
and thus we have from Eqs.~(\ref{delta:decom}), (\ref{Dplus:soln})
and~(\ref{C1early}) the first-order solution 
\beq
\label{delta1final}
\delta^{(1)} = \left[ f_1(\Omega_m)+\frac32\Omega_m \right]^{-1} 
\frac{\nabla^2 \Rc}{\Hs^2} \,. 
\eeq

For the purposes of the subsequent analysis, and since we are working
with a single degree of freedom, let us relate all our variables with
the curvature perturbation $\Rc$. {}From the
continuity equation \eqref{Ch6:eq:consfirst}, and using
Eqs.~\eqref{delta:decom} and \eqref{Ch6:eq:fm} to eliminate $D_+'$, we
find   
\begin{eqnarray}
\vartheta^{(1)} 
\label{theta1:Rc1}
= -C_1(\bx) D_+'(\tau)
= - \left[ \frac{f_1(\Omega_m)}{f_1(\Omega_m)+(3/2)\Omega_m} \right] 
\frac{\nabla^{2}\Rc}{\Hs}.
\end{eqnarray}

\noindent Integrating \eref{theta:trace} gives
\begin{eqnarray}
\psi^{(1)} 
= \frac13 C_1(\bx) D_+(\tau)  + \Rc(\bx)
= \frac13 \left[ f_1(\Omega_m)+\frac32\Omega_m \right]^{-1} 
\frac{\nabla^2 \Rc}{\Hs^2}  + \Rc\,. 
\end{eqnarray}

\noindent where the constant of integration is set from the definition
of $\Rc$ in \eref{R1c:expansion}. We choose initial spatial
coordinates such that on large scales/early times $\psi^{(1)}\sim
\Rc$, consistent with the separate universe approach, where the metric
approaches a manifestly FRW metric at large scales
\citep{Lyth:1984gv,Wands:2000dp}.    

The expansion of $\Rc$ in Eqs.~\eqref{R1:expansion} and
\eqref{R1:Rc1} makes trivial the determination of $\nabla^2\chi^{(1)}$ in
terms of this variable, 
\begin{eqnarray}
\chi^{(1)} 
\label{chi1:Rc1}
= -2 \nabla^{-2} C_1(\bx) D_+(\tau) 
= -2 \left[ f_1(\Omega_m)+\frac32\Omega_m \right]^{-1} \frac{\Rc}{\Hs^2}\,,
\end{eqnarray}

\noindent $\vartheta^{(1)}{}^i_j$ then follows from \eref{theta1:chi1}. 
Having assumed a purely growing mode, at early times, $\tau \to 0$, 
the density contrast, inhomogeneous expansion and shear 
are suppressed and the only surviving perturbation is the primordial
curvature perturbation, $\psi^{(1)}\to \Rc$.  

{In this paper we are primarily interested in connecting initial
conditions at the beginning of the matter-dominated era with
primordial fluctuations. In this case, although the solution
\eref{delta1final} is general, it is of practical use only if one has
a solution for ${f_1}$, such as the approximation \eref{f1:Omegam}. On
the other hand, our first integral \eref{first:intdelta} provides a
straightforward method to obtain an explicit solution for the growing
mode. Indeed the latter, as mentioned above, can be obtained as the
particular solution of \eref{first:intdelta}. Rewriting this equation as 
\beq
\label{delta:a}
{\frac{d \delta^{(1)}}{d a} +\frac32 \frac{\Omega_m}{a} \delta^{(1)} =
\frac{\Omega_m}{4 \Hs_0^2\Omega_{m0}}\Rs^{(1)},}
\eeq

\noindent and using standard methods for solving first order
inhomogeneous equations we obtain the particular solution
\beq
\label{exact:delta+}
{\delta^{(1)}(a) = D_- \frac{\Rs^{(1)}}{4\Hs_0^2\Omega_{m0}} \int_0^{a}
\frac{\Omega_m}{D_-} \,da = \nabla^2 \Rc \left(\frac{\Hs}{a}\right)
\int_0^{a} \frac{da}{\Hs^3(a)},} 
\eeq 

\noindent where we used the homogeneous solution of \eref{delta:a},
i.e.\ the decaying mode ${D_-\propto \Hs / a}$. 

To summarise, Eqs.~\eqref{delta1final} and \eqref{exact:delta+} give
the growing mode of the linear density perturbation, directly in terms of the
early universe curvature fluctuation ${\Rc}$, with no arbitrary
constants.}
%
\subsection{Second-order solution}\label{Ch6:sec:second}
%
We shall now derive the second order differential equation for
$\delta^{(2)}$ following the method developed in the first-order
analysis. Our method does not require solving for
the second-order metric variables, which is a simplification to other
works \citep{Bar10}. We start 
with the second-order perturbative expansion of the continuity equation 
\eref{Ch6:eq:conserv}, that is  
\beq 
\label{Ch6:eq:consdel2}
{\delta^{(2)}}' + \vartheta^{(2)} = - 2{\delta^{(1)}}\vartheta^{(1)}.
\eeq

\noindent The Raychaudhuri equation, \eref{Ch6:eq:Ray}, expanded at
second order is 
\beq
{\vartheta^{(2)}}'+\Hs\vartheta^{(2)}
+\frac{3}{2}\Hs^2\Omega_m\delta^{(2)}= - 2{\vartheta^{(1)i}}_j{\vartheta^{(1)j}}_i.
\eeq 

\noindent As before, we combine the last two equations to derive the
evolution equation for $\delta^{(2)}$, finding
\beq 
\label{Ch6:eq:odedelta1} 
{\delta^{(2)}}''+
\Hs{\delta^{(2)}}'-\frac{3}{2}\Hs^2\Omega_m\delta^{(2)} = 
-2{\delta^{(1)}}'\vartheta^{(1)}-2\delta^{(1)}{\vartheta^{(1)}}'-2\Hs
\delta^{(1)}\vartheta^{(1)}+ 2{\vartheta^{(1)i}}_j{\vartheta^{(1)j}}_i\,.
\eeq

\noindent We can eliminate $\vartheta^{(1)}$ and $\vartheta^{(1)'}$
with the aid of
Eqs.~\eqref{Ch6:eq:consfirst}~and~\eqref{Ch6:eq:Rayfirst}. We also use
Eq.~\eqref{theta1:chi1} and write
\beq
{\delta^{(2)}}'' + \Hs{\delta^{(2)}}' -
\frac{3}{2}\Hs^2\Omega_m\delta^{(2)}
= 2(\delta^{(1)}{}'){}^2 + 3 \Hs^2 \Omega_m (\delta^{(1)}){}^2+
\frac12{\partial^i\partial_j{\chi^{(1)}}'}{\partial^j\partial_i
  {\chi^{(1)}}'}\,.
\label{evol:reduced1}
\eeq

\noindent The final form of the evolution equation is found using the
first-order solutions for $\delta^{(1)}$ and $\chi^{(1)}$ presented in the previous section, that is
\beq
\label{evol:reduced2}
{\delta^{(2)}}'' + \Hs{\delta^{(2)}}' -
\frac{3}{2}\Hs^2\Omega_m\delta^{(2)} =
\left[2 f_1^2 + 3 \Omega_m + 2\Sigma f_1^2\right] C_1^2 D_+^2(\tau) ,
\eeq

\noindent where we introduce the shape coefficient
\beq
\label{sigma:def}
\Sigma \equiv \frac{\vartheta^i_j \vartheta^j_i}{\vartheta^2} =
\frac{\partial_i\partial_j \Rc \partial^i\partial^j 
  \Rc}{(\nabla^2 \Rc)^2}.
\eeq 

Instead of directly solving the evolution \eref{evol:reduced2}, let us
look at a first integral of the evolution in the energy constraint
\eref{Ch6:eq:evol}, as we did in our analysis at first order. Expanded
at second order this constraint is      
\beq 
\label{Ch6:eq:GRconstr}
4\Hs\vartheta^{(2)} - 6\Hs^2\Omega_m\delta^{(2)}+\Rs^{(2)} 
= 2 \vartheta^{(1)i}_j\vartheta^{(1)j}_i - 2 \vartheta^{(1)2}, 
\eeq 

\noindent where the second-order Ricci scalar, $\Rs^{(2)}$, is given in
terms of metric perturbations by 
\citep{MatMolBru98},   
\beqa  
\label{Ch6:eq:threericdeco}
\frac12\Rs^{(2)}&=& 2\nabla^2\left[
  \psi^{(2)}+\frac16\nabla^2\chi^{(2)}\right]+ 6\left( \nabla
\psi^{(1)}\right)^2\nonumber\\ 
&+&16\psi^{(1)}\nabla^2\psi^{(1)} +
4\psi^{(1)}\partial_l\partial_j\chi^{(1)lj}
-2\partial_j\partial_k\psi^{(1)}\chi^{(1)jk} \nonumber \\ &+&
\chi^{(1)jk}\nabla^2\chi^{(1)}_{jk}
-2\chi^{(1)jk}\partial_l\partial_k{\chi^{(1)l}}_j-
\partial_l\chi^{(1)lk}\partial_j{\chi^{(1)j}}_k \nonumber\\ &
+&\frac34\partial_k\chi^{(1)lj}\partial^k\chi^{(1)}_{lj}
-\frac12 \partial_k\chi^{(1)lj}\partial_l{\chi^{(1)k}}_j,  
\eeqa

\noindent an equation purely dictated by geometry. 

Combining \eref{Ch6:eq:GRconstr} with the continuity
\eref{Ch6:eq:consdel2} to eliminate $\vartheta^{(2)}$, we obtain:
\beq
\label{first:integral2}
 4 \Hs {\delta^{(2)}}' + 6\Hs^2\Omega_m\delta^{(2)}  - \Rs^{(2)}
 = 2 \vartheta^{(1)2}- 2 \vartheta^{(1)i}_j\vartheta^{(1)j}_i - 
 8\Hs \delta^{(1)}\vartheta^{(1)}\,.   
\eeq

As we saw from the first-order analysis
Eqs.~\eqref{first:intdelta}~and~\eqref{R:const}, the equation for 
$\delta^{(2)}$ is coupled with an equation for $\Rs^{(2)}$. At first
order, $\Rs^{(1)}$ is conserved and \eref{first:intdelta} is a first integral of
\eref{Ch6:eq:deltadd2}. At second order, the time derivative of
\eref{Ch6:eq:GRconstr} can be reduced to  
\beq
\label{R2:prime}
{\Rs^{(2)}}' = - 4 {\vartheta^{(1)}}^{i}_{~j}{\Rs^{(1)}}^j_{~i} = - 2
\left[\partial^i \partial_j \chi^{(1)}{}' \partial^j\partial_i \Rc + 
\nabla^2 \chi^{(1)}{}' \nabla^2 \Rc \right].
\eeq
One can also derive \eref{R2:prime} by taking the time derivative of  \eref{Ch6:eq:threericdeco} for $\Rs^{(2)}$ and using the momentum constraint \eref{Ch6:eq:momcon}.

Given the correspondence between the left-hand sides of
Eqs.~\eqref{first:integral2} and \eqref{R2:prime} to their first-order
equivalents, Eqs.~(\ref{first:intdelta}) and~(\ref{R:const}), we solve
the coupled system of these equations by separating the solution as
follows: 
\beq
\delta^{(2)} = \delta^{(2)}_h + \delta^{(2)}_p, \qquad \Rs^{(2)} =
      {\Rs^{(2)}_h} + {\Rs^{(2)}_p}, 
\label{ricci2:split}
\eeq

\noindent where $\Rs_h^{(2)}$ is the constant solution to the
homogeneous part of \eref{R2:prime} and $\Rs_p^{(2)}$ is the
particular solution, and $\delta_h^{(2)}$ is the solution of
\eref{first:integral2} that is generated by assuming $\Rs_h^{(2)}$ as
the only source term, see Eqs.~\eqref{hom:delta2}.
%
\subsubsection{The particular solution}
%
The particular solution of the inhomogeneous system is obtained by
integrating \eqref{R2:prime} directly, yielding    
\beq
\label{R2:part}
{\Rs^{(2)}_p}  = - 2\left[\partial^i \partial_j \chi^{(1)}
  \partial^j\partial_i \Rc + \nabla^2 \chi^{(1)}\nabla^2 \Rc \right].
\eeq

\noindent This contributes to the non-linear driving terms in
\eref{first:integral2}, and thus the particular part of the solution. Using \eref{chi1:Rc1} for $\chi^{(1)}$, we can write \eref{first:integral2} as
\begin{equation}
{4 \Hs \delta_p^{(2)}{}' + 6 \Hs^2 \Omega_m \delta^{(2)}_p = (\nabla^2 \Rc )^2 S(\tau ,\Sigma )\,,}
\label{delp2:f2}
\end{equation}

\noindent {where {$S(\tau,\Sigma)$} is a function of time and the shape coefficient {$\Sigma$} introduced in \eref{sigma:def},}
\beq
{S(\tau,\Sigma ) = \frac{2\left[(2 f_1 + 3\Omega_m )(1 + \Sigma ) + f_1^2 (1 - \Sigma ) + 4 f_1 \right]}{\left(f_1 + \frac{3}{2}\Omega_m\right)^2\Hs^2} \,.}
\eeq

{Equation \eqref{delp2:f2} is an inhomogeneous linear
  ordinary differential equation, whose solution has a standard
  integral form in terms of the source term {$(\nabla^2 R_c)^2
  S(\tau,\Sigma)$}. Given the factorised form of this source term, we
  can then write} 
\beq
 \label{delta2p}
\delta^{(2)}_p = P(\bx) D_{2 +}(\tau,\Sigma),
\eeq

\noindent { with {$D_{2 +}(\tau_0,\Sigma) = 1$}, and thus
  {$P(\bx) \equiv \delta_p^{(2)}(\bx,\tau_0)$}. In analogy to the
  first-order case \eref{Ch6:eq:fm}, we can now define}
\beq
{f_2 \equiv \frac{1}{2}\frac{D_{2+}'}{\Hs D_{2+}}\,.}
\eeq

At early times, during matter domination ($\Omega_m = 1$), we have the
solution $D_{2+} \propto (D_{1+})^2 \propto a^2$, hence
$f_2(1,\Sigma) = 1$. 
In a universe dominated by dust, using \eref{delta2p} in
\eref{delp2:f2} gives 
\beq
P(\bx) = \frac{2(5 + 2 \Sigma)}{7} \left(\frac{D_{+
    IN}^2}{D_{2+ IN}}\right)  C_1^2(\bx) .
\label{early:P}
\eeq
More generally, \eref{delp2:f2} can be written as 
\beq
\left[ \frac{4 f_2 + 3 \Omega_m}{7} \right] 
\frac{D_{2+}(\tau,\Sigma)}{D_{2+ IN}} = 
\left[ \frac{(2 f_1 + 3\Omega_m)(1 + \Sigma) + f_1^2 (1 - \Sigma) 
    + 4 f_1 }{2(5+2\Sigma)} \right]
\frac{D_+^2(\tau)}{D_{+ IN}^2}
\,.
\label{part:soln2}
\eeq
Thus the second order particular solution \eref{delta2p} can be written as
\beq
\delta^{(2)}_p =
\left[ \frac{(2 f_1 + 3\Omega_m)(1 + \Sigma) + f_1^2 (1 - \Sigma) + 4 f_1 }{4 f_2 + 3 \Omega_m} \right]
\left( \delta^{(1)} \right)^2
\,,
\eeq
or, using \eref{delta1final} and substituting in for the shape coefficient \eref{sigma:def},
\beq
 \label{delta2pfinal}
\delta^{(2)}_p =
\frac{\left( 6f_1+f_1^2+3\Omega_m \right) \left( \nabla^2\Rc \right)^2 + \left( 2f_1-f_1^2+3\Omega_m \right) \partial^i\partial_j\Rc \partial_i\partial^j\Rc }{\left( 4 f_2 + 3 \Omega_m \right) \left( f_1+\frac32\Omega_m \right)^2 \Hs^4}
\,.
\eeq

For $\Omega_m \simeq 1$, we can set $f_2 (\Omega_m,\Sigma)\simeq
\Omega^p_m $ (as we did at first order), where $p=p(\Sigma)$ can be
determined by taking the time derivative of the logarithm of \eref{part:soln2},
\beq 
\frac{4p + 3}{7}\Omega_m' + 2 \Hs \Omega^{p}_m \simeq
\frac{(48/11)  +  3 (1 + \Sigma)}{10 + 4 \Sigma}\Omega_m' 
+ 2\Hs\Omega_m^{6/11},
\label{p:omega}
\eeq 

\noindent where we have used \eref{f1:Omegam} for $f_1$. The time dependence of
$\Omega_m$, at first order in $1 - \Omega_m$, is given from the
background \eref{Omega:def};  $\Omega_m'\simeq 3 \Hs (\Omega_m -
1)$. Thus, we obtain 
\beq
\label{p:explicit}
26\, p = -\frac{15}{11} + \frac{21}{5+ 2\Sigma} \left[\frac{24}{11} +
  \frac32 (1 + \Sigma)\right].
\eeq

For the special case of planar symmetry, the shape
coefficient \eref{sigma:def} has the value $\Sigma = 1$. In this case
\eref{part:soln2} directly shows that $f_{2} (\Omega_m,1)=
f_1(\Omega_m)$ and hence $D_{2+} \propto (D_{1+})^2$. The particular
solution \eref{delta2pfinal} then reduces to a remarkably simple form 
\beq
\delta^{(2)}_p = 2\left( \delta^{(1)} \right)^2 \,.
\eeq

\noindent The planar case $\Sigma=1$ describes an exact solution
leading to the formation of a pancake during gravitational
collapse. This exact solution is an important case because it plays
the role of an attractor solution in the Zel'dovich approximation in
the Newtonian framework
\citep{Zeldovich:1969sbxb,buchert:92,maartens-bruni}.   

\subsubsection{The homogeneous solution}
For the homogeneous part of the solution we solve the same system of
coupled equations, \eqref{first:intdelta}~and~\eqref{R:const}, as at
first order,  
\begin{align}
\label{hom:delta2}
4 \Hs \delta^{(2)}_h{}' + 6 \Hs^2 \Omega_m \delta^{(2)}_h - 
\Rs_{h}^{(2)} &= 0, \\
\Rs_h^{(2)}{}' &= 0.  
\end{align}

\noindent Therefore $\Rs_h^{(2)} = {\rm const.}$, and the homogeneous
growing mode solution is given by $\Rs_h^{(2)}\neq 0$, 
\beq
 \label{delta2h}
\delta_h^{(2)}(\tau,\bx ) = C_2(\bx )D_+ (\tau),
\eeq

\noindent where $D_+(\tau)$ is the linear growth factor in
\eref{expressD+} or \eref{exact:delta+} and  
\begin{eqnarray}
\label{def:C2}
C_2({\bx})
&=&  \frac{\Rs^{(2)}_h({\bx}) }{10 \Hs_{IN}^2 D_{+IN}} \,,
\end{eqnarray}

\noindent {in complete analogy with \eref{C1early}.} The second-order
constant of integration, $\Rs_h^{(2)}$, is derived in terms of
second-order metric variables by subtracting
the particular solution, \eref{R2:part}, from the expression for
$\Rs^{(2)}$ given in \eref{Ch6:eq:threericdeco}, {which yields} 
\beqa
\frac12 \Rs^{(2)}_h &=& 2 \nabla^2\left[\psi^{(2)} + \frac16\nabla^2
  \chi^{(2)}\right] + 16 \Rc \nabla^2 \Rc + 6 \partial^k \Rc
\partial_k \Rc 
\label{R2:homogeneous}\\
&&
- \left[ 2 \partial^k \nabla^2 \chi^{(1)} \partial_k \Rc + \partial^i \partial_j \chi^{(1)}
  \partial^j\partial_i \Rc + \nabla^2 \chi^{(1)}\nabla^2 \Rc \right]
\nonumber\\
&&
 + \frac14 \left[\partial^i \partial^j\partial^k \chi^{(1)} 
\partial_i \partial_j\partial_k \chi^{(1)} - 
\partial^k \nabla^2 \chi^{(1)}\partial_k \nabla^2 \chi^{(1)}\right].
\nonumber
\eeqa

\subsubsection{The complete solution}

The general growing-mode solution for the second-order density
perturbation \eref{ricci2:split} can thus be written in terms of
\eref{delta2p} and \eref{delta2h} as 
\beqa
\delta^{(2)} 
&=& P(\bx) D_{2 +}(\tau,\Sigma) + C_2(\bx ) D_+ (\tau) \nonumber\\
&=& \frac{2(5 + 2 \Sigma)D_{+ IN}^2}{7D_{2+ IN}} C_1^2(\bx) D_{2 +}(\tau,\Sigma) 
+ \frac{1}{10 \Hs_{IN}^2 D_{+IN}} \Rs^{(2)}_h({\bx}) D_+ (\tau).
\label{delta2:complete}
\eeqa

In principle we are free to set the constant $\Rs^{(2)}_h$ at any
time. In practice we wish to determine $\Rs^{(2)}_h$ in terms of the
primordial metric perturbations, using \eref{R2:homogeneous}. We can
do this on scales which lie outside the horizon at the start of the
matter-dominated era. 
This limits the range of scales for which our subsequent analysis is
valid; only those modes larger than the horizon at the equality of
radiation and matter will obey the following initial conditions (this
corresponds to wavenumbers $k \gtrsim (90\, \mathrm{Mpc})^{-1} h$
\citep{lyth:liddle}). On smaller scales we could use the output of
numerical second-order Einstein-Boltzmann codes,
e.g. \citep{pettinari:13}, and construct the complete expression in
\eref{R2:homogeneous} on all scales at the start of the
matter-dominated era.  

The primordial curvature perturbation is commonly given in terms of
the non-linear variable $\zeta$ \citep{MalWan09}, such that 
\beq
\exp(2\zeta) = 1 - 2 \left[\psi + \frac16 \nabla^2\chi \right]\,.
\label{exp:zeta}
\eeq

\noindent This extends the definition in \eref{zeta:def} to non-linear
orders. $\zeta$ is constant for adiabatic density perturbations in the
long-wavelength limit
\citep{MalWan04,Lyth:2004gb,Langlois:2005ii}. Since at the end of
inflation all scales lie far outside the horizon, this variable is
commonly used to describe the primordial curvature perturbation beyond
linear order.  
In particular it is used to define primordial non-Gaussianity, and the non-linearity parameter $f_{\rm NL}$ \citep{LytRod05}. 
For local-type non-Gaussianity, 
we have the second-order expansion 
\beq
\label{zeta:fnl}
\zeta_{IN} = {\zeta^{(1)}}_{IN} +  \frac35 f_{\rm  NL}{\zeta^{(1)}}_{IN}^2, 
\eeq

\noindent where ${\zeta^{(1)}}_{IN}({\bx})$ is a first-order Gaussian
random field due to quantum vacuum fluctuations during inflation. A
Gaussian distribution of primordial perturbations from inflation 
then corresponds to $f_{\rm NL}=0$ \citep{Maldacena:2002vr}.

The non-linear definition \eref{exp:zeta}, expanded at second order
and using \eref{zeta:fnl}, 
yields  
\beq 
\label{psi:zeta}
\psi^{(2)}_{IN} + \frac16 \nabla^2\chi^{(2)}_{IN} 
= - \left(\frac65 f_{\rm NL} + 2\right) {\zeta^{(1)}}_{IN}^2 
= - \left(\frac65 f_{\rm NL} + 2\right) {\Rc}^2. 
\eeq
Note that we only expect the primordial perturbation $\zeta$ to remain
constant to leading order in a gradient expansion. Therefore we can
only use \eref{psi:zeta} to set the initial conditions on large scales
and early times in the matter era, $k^2\ll \Hs_{IN}^2$. Thus,
combining \eref{psi:zeta} and \eref{R2:homogeneous}, we set the
initial conditions for super-horizon scales at the start of the
matter-dominated era 
\beq
\frac12 \Rs^{(2)}_h 
 \simeq - 2 
 \nabla^2 \left[ \left( \frac65 f_{\rm NL} + 2 \right) {\Rc}^2 \right]
  + 16 \Rc \nabla^2 \Rc + 6 \partial^k \Rc
\partial_k \Rc  \,.
\eeq

\noindent \eref{chi1:Rc1} shows explicitly that $\chi^{(1)}\propto
\Rc / \Hs^2$, therefore, the terms involving $\chi^{(1)}$ (the second
and third lines) in \eref{R2:homogeneous} are subdominant for modes
larger than the initial horizon scale, $k^2 \ll \Hs_{IN}^2$. In this
limit we can write the homogeneous part of the second-order density
perturbation solution \eref{delta2h} as  
\beq
\delta^{(2)}_{h} = - \frac{12}{5\Hs^2}\left[f_1(\Omega_{m})+\frac32
  \Omega_{m} \right]^{-1} \Bigg\{ \left(f_{\rm NL} -
  \frac53\right) \Rc\nabla^2\Rc + \left( f_{\rm NL} + \frac{5}{12} \right)
  \partial^j\Rc \partial_j\Rc \Bigg\}
\,.
\label{soln:delta2h}  
\eeq 

\noindent This homogeneous solution illustrates how primordial
non-Gaussianity is transferred to the matter perturbations in a manner
consistent with GR. Additionally it shows how GR itself leads to
non-linear constraint equations which contributes to the initial
non-Gaussianity of the matter density field, even if $f_{\rm NL}=0$.  

Finally, we can give the complete solution, \eref{delta2:complete},
combining the homogeneous solution, \eref{soln:delta2h}, and the
particular solution, \eref{delta2pfinal}. At leading order in a
gradient expansion, we obtain
\beqa  
 \label{delta2final}
\delta^{(2)}(\bx,\tau)&=& -\frac{12}{5\Hs^2} 
\left[ f_1+\frac32 \Omega_{m} \right]^{-1} 
\Bigg\{  \left( f_{\rm NL} - \frac53 \right) \Rc\nabla^2\Rc +
\left( f_{\rm NL} + \frac{5}{12} \right) \partial^j\Rc \partial_j\Rc 
 \Bigg\}\,+
 \nonumber\\ 
&\,&
\frac{\left( 6f_1+f_1^2+3\Omega_m \right) \left( \nabla^2\Rc \right)^2
  + \left( 2f_1-f_1^2+3\Omega_m \right) \partial^i\partial_j\Rc
  \partial_i\partial^j\Rc }{\left( 4 f_2 + 3 \Omega_m \right) \left(
  f_1+\frac32\Omega_m \right)^2 \Hs^4}\,, 
\eeqa

\noindent with the first line representing the primordial
non-Gaussianity and GR correction, which dominate at large
scales, and the last line corresponding to the growing Newtonian
solution, which dominates on small scales. 

\eref{delta2final} agrees with the second-order density perturbation
in the synchronous-comoving gauge presented in Eq.~(7) of \cite{Bar05}
in the matter-dominated limit, $\Omega_m=1$. On the other hand, our
solution is consistent with Eq.~(5.1) of
\cite{tomita:67}, Eq.~(4.39) of \cite{MatMolBru98} and Eq.~(40) of
\cite{Hwang:2012bi}, in the matter-dominated era for the  particular 
choice of primordial non-Gaussianity parameter  $f_{\rm NL}=-5/3$ (the
equivalence with \cite{Hwang:2012bi} is explicit when 
the solution is written in Eulerian spatial coordinates, see
Appendix). We can also compare the second-order solution in a
$\Lambda$CDM-cosmology with that presented  by \cite{Tomita:2005et},
where Eq.~(2.22) is consistent with our result for the particular case
{$f_{\rm NL}=-5/3$}. We finally comment on the result of
\cite{Bar10}. Their solution for the second-order density perturbation
is written in terms of the initial potential, $\Phi_{IN}=(3/5)\Rc$,
and the linear growth  
factor 
\eref{expressD+}  
\beq
D_+(\tau) = \left(
\frac{f_1(\Omega_{m0})+\frac32\Omega_{m0}}{f_1+\frac32\Omega_m }
\right)\frac{\Hs^2_0}{\Hs^2} 
\,,
\eeq 
under the approximation $D_{2+}\propto D_+^2$, which we have seen only
strictly holds during matter-domination ($\Omega_m=1$) or for the
planar case $\Sigma=1$. In this approximation, we have 
\beqa  
\delta^{(2)}(\bx,\tau)&=& -\frac{20}{3\Hs^2_0} 
\left[ f_1(\Omega_{m0})+\frac32 \Omega_{m0} \right]^{-1} 
\Bigg\{  \left( f_{\rm NL} - \frac53 \right) \Phi_{IN}\nabla^2\Phi_{IN} +
\left( f_{\rm NL} + \frac{5}{12} \right) \partial^j\Phi_{IN} \partial_j\Phi_{IN}
 \Bigg\} D_+(\tau)
 \nonumber\\ 
&& +\,
\frac{50}{63\Hs^4_0} 
\left[ f_1(\Omega_{m0})+\frac32 \Omega_{m0} \right]^{-2} 
\left[ 5 \left( \nabla^2\Phi_{IN} \right)^2 + 2
  \partial^i\partial_j\Phi_{IN} \partial_i\partial^j\Phi_{IN} \right] 
 D_+^2(\tau) \,.
\eeqa
remembering that in fact $\partial^i\partial_j\Rc
\partial_i\partial^j\Rc=( \nabla^2\Rc)^2$ when $\Sigma=1$. 
This expression agrees with the solution presented in Eq.~(45) of
\cite{Bar10} if we adopt their non-linearity parameter $a_{\rm
  NL}=1+(3/5)f_{\rm NL}$.

\section{Relation to Newtonian results}\label{Ch6:sec:NGR} 
%
\subsection{Equivalence of non-linear evolution equations}
\label{Ch6:subsec:equiv}
In the Newtonian treatment, the continuity equation in Eulerian
coordinates is  
\beq 
\label{Ch6:eq:dervis} 
\frac{\partial
  \delta_{\rm N}}{\partial \tau} + \nabla \cdot \left[(1+\delta_{\rm
    N}){\bf v}\right] = 0\,,  
\eeq 

\noindent where $\tau$ denotes conformal time, $\mathbf{v} =
\mathbf{x}'$ and $\bx$ is the background comoving coordinate. The
Euler equation, which dictates the flow evolution, is 
\beq 
 \label{Euler}
\frac{\partial {\bf v}}{\partial \tau}+\Hs{\bf v}
+\left( {\bf v}\cdot\nabla\right){\bf v} + \nabla \varphi = 0\,. 
\eeq

\noindent In the Newtonian theory, the system is closed by the
Poisson equation
\beq
\label{newtonian:poisson}
\nabla^2 \varphi =  \frac{3}{2}\Hs^2\Omega_m \delta_N\,.
\eeq

\noindent Unlike the relativistic Poisson equation \eqref{Ch6:eq:Poi}, 
which is only valid at first order, in the Newtonian case this
equation is exact.  

Combining the divergence of \eref{Euler} with the Poisson equation gives 
\beq 
\label{rai:newtonian}
\frac{\partial\left(\nabla\cdot {\bf v} \right)}{\partial \tau}+\Hs
\nabla \cdot{\bf v} +\nabla\cdot\left( {\bf v}\cdot\nabla\right){\bf
  v}+\frac{3}{2}\Hs^2\Omega_m\delta_{\rm N}=0\,.
\eeq

\noindent One can define a Newtonian deformation tensor as
$\vartheta_{\rm N}^{ij} = \partial^i v^j$, with trace $\vartheta_{\rm
  N}$. For an irrotational fluid there is a velocity potential ${\bf
  v}=\partial^i v$, and we can expand    
\beq 
\label{identity}
\nabla \cdot \left( {\bf v} \cdot \nabla\right) {\bf v}  =
\partial_i\left( v^j\partial_j\right)v^i=\partial_i \partial^j v
\partial_j\partial^i v+\partial^j v \partial_j\nabla^2 v\, = 
\vartheta^i_{{\rm N} j} \vartheta^j_{{\rm N}i }  + \partial^j
v\partial_j\vartheta_{\rm N}.  
\eeq 

\noindent Introducing the Lagrangian time derivative, related to 
the Eulerian derivative through 
\beq
\frac{d}{d\tau} = \frac{\partial }{\partial \tau} + {\bf v}\cdot
\nabla\,,
\label{lagrangian:derivative}
\eeq

\noindent and using \eref{identity}  we can write
\eqref{Ch6:eq:dervis} and \eref{rai:newtonian} as 
\begin{eqnarray}
\label{cont:newton1}
&\,&\frac{d\delta_{\rm N}}{d \tau} + (1+\delta_{\rm N})\vartheta_{\rm N} = 0\,,\\
\label{rai:newton1} 
&\,&\frac{d\vartheta_{\rm N}}{d\tau}+\Hs \vartheta_{\rm N} +
\vartheta^i_{{\rm N} j} \vartheta^j_{{\rm N}i} +
\frac{3}{2}\Hs^2\Omega_m\delta_{\rm N}=0\,,  
\end{eqnarray}

\noindent i.e., the continuity equation and the Newtonian Raychaudhuri
equation in Lagrangian form \citep{Pee80}. 
The formal equivalence of these Newtonian evolution equations with
their relativistic counterpart Eqs.~\eqref{nonlinear:continuity} and
\eqref{Ch6:eq:Ray} is evident, as long as one keeps in mind that a
partial derivative with respect to the synchronous-comoving time $\tau$ in the
relativistic case, corresponds to the convective Lagrangian derivative in the
Newtonian case. The difference remains in the constraint equations
\citep{MatPanSae94}, i.e. the energy constraint \eref{Ch6:eq:evol} and the
momentum constraint \eref{Ch6:eq:momcon} in the relativistic case,
versus the Poisson equation \eqref{newtonian:poisson}. At first order,
the equivalence of the dynamics is complete because the first-order
Poisson equation, \eref{Ch6:eq:Poi}, combines both relativistic
constraint equations.  

The equivalence of the equations allow us to establish a
correspondence between Newtonian and relativistic variables, which we
summarise as 
\beqa
{\rm Newtonian\ Lagrangian}\quad &\leftrightarrow& 
\quad {\rm Relativistic\ comoving} \nonumber\\
\frac{d}{d\tau} \quad &\leftrightarrow& \quad   
\frac{\partial}{\partial\tau} \nonumber\\
\partial^i{\rm v}_{{\rm }j} \quad &\leftrightarrow& \quad  
\vartheta^i_{~j} \nonumber\\
\delta_{\rm N} \quad &\leftrightarrow& \quad  \delta\nonumber\\
\varphi^{(1)}_{IN}
 \quad &\leftrightarrow& \quad  \frac{3}{5}{\Rc}\nonumber
\eeqa

\noindent While the correspondence between deformation tensors and between
density contrast is exact, that between metric potentials is only
valid at first order in the matter dominated era, where
$\Phi_{IN}= \varphi^{(1)}_{IN}$. In addition, we note that the
interpretation of $\vartheta^i_j$ is gauge-dependent; it is only in
our synchronous-comoving gauge that it coincides with the extrinsic
curvature, i.e., the deformation of the space slices. 
In general, in other gauges $\vartheta_{i}^j$ 
will contain a Newtonian term $\partial_i v^j$ as well as 
post-Newtonian contributions from the metric 
\citep{BruDunEll92,milillo}.

\subsection{Initial conditions for numerical simulations}
\label{Ch6:subsec:spec} 
%
%
Our results suggest that numerical simulations of large-scale
structure formation based on Newtonian evolution equations can
describe general relativistic evolution, even on super-horizon scales,
in the case of irrotational flow in a $\Lambda$CDM cosmology, where
the above dictionary should be used to interpret Newtonian variables
in terms of their ``true'' relativistic meaning.  
In numerical codes, the first stage of the evolution of matter
fluctuations is computed by solving the Newtonian equations of
hydrodynamics at non-linear orders in a perturbative expansion. 
However, we have seen that care needs to be taken when setting initial
conditions such that they respect the non-linear constraint equations
of general relativity. 

In Newtonian gravity, the Poisson equation is
valid at all orders of perturbation. This is customarily used to
account for primordial non-Gaussianity of local type
(see e.g. \cite{Dalal:2007cu,Scoccimarro:2011pz}) through the expansion 
\beq
\delta_{IN} =  \frac{2}{3\Hs^2_{IN}} 
\nabla^2 \varphi_{IN} = 
 \frac{2}{3\Hs^2_{IN}} 
 \nabla^2 \left[\Phi_{IN} - f_{\rm NL} \Phi_{IN}^2 \right]\,, 
\label{newtonian:poisson2linear}
\eeq

\noindent where $\Phi_{IN}$ is a Gaussian first-order
potential\footnote{Often in the Newtonian context a different sign
  convention is used \citep{Wands:2010af}.}. 
The above constraint, or its Fourier counterpart, is
imposed at some initial redshift, usually of the order $z(\tau_{IN}) \sim
10^2$, in the matter-dominated universe. 

In the GR framework, the above equivalence is incomplete since
the Poisson equation \eqref{Ch6:eq:Poi} is valid only for linear perturbations 
\citep{Wands:2009ex}. The above initial 
condition (\ref{newtonian:poisson2linear}) disregards non-linear GR
contributions. The result in \eref{delta2final} provides the correct
constraint on large scales, consistent with GR. Therefore the initial
condition, at second order, 
in terms of the Gaussian potential $\Phi_{IN}=(3/5)\Rc$ yields the result 
\beq  
 \label{gr:poisson2}
\frac12 \delta^{(2)}_{IN} = -\frac{4}{3\Hs^2_{IN}} 
\left[  \left( f_{\rm NL} - \frac53 \right) \Phi_{IN}\nabla^2\Phi_{IN} +
\left( f_{\rm NL} + \frac{5}{12} \right) \partial^j\Phi_{IN} \partial_j\Phi_{IN}
 \right] \,.
\eeq
Note that we have omitted here the particular part of the solution in
\eref{delta2final} since that is sub-dominant on large-scales and at
early times, and because it is generated by the subsequent Newtonian
evolution. Initial conditions in the comoving-Poisson gauge have
been presented as a solution for $\delta^{(2)}$ \citep{Bar10} or in terms
of the constraint equations \citep{Hidalgo:2013mba}. 

In numerical simulations, the initial conditions are set by
approximate solutions 
to the non-linear equations. The codes fulfilling this task, dubbed
initial condition generators, follow most commonly the Zel'dovich
approximation \citep{Zeldovich:1969sbxb} or the more accurate second
order solution in Lagrangian coordinates (2LPT)
\citep[e.g.][]{Scoccimarro:1997gr}. Our prescription for initial
conditions sets a precedent to make the initial condition generators
consistent with general relativity. 

We show in the appendix that the second-order density perturbation
\eref{gr:poisson2} includes a contribution in the squeezed limit
equivalent to a primordial non-Gaussianity parameter $f_{\rm NL} = - 5 / 3$. 
Equivalently, the final term in \eref{gr:poisson2} may be neglected
when considering the statistics of virialised objects, corresponding
to peaks of the matter density field. Around these maxima one could
argue that $\nabla \Phi \approx 0$ \citep{Dalal:2007cu}. Therefore,
comparing with \eref{newtonian:poisson2linear}, the
relativistic corrections yield an effective contribution of magnitude $f_{\rm NL}
= - 5 / 3$.   
%
%
%
\section{Discussion}\label{Ch6:sec:sum}
%
Our starting point is the system of coupled non-linear evolution
equations for the inhomogeneous expansion, $\vartheta$, and density
contrast, $\delta$, for an irrotational flow in general relativity in
the synchronous-comoving gauge, adopting the spirit of the fluid-flow
approach to cosmological perturbations. We have seen that these
evolution equations correspond to the Newtonian evolution equations
where we identify the comoving density with the Lagrangian density,
and the expansion with the divergence of the Eulerian velocity, $v$. 

At first order in perturbative expansion, the correspondence
$\vartheta \to \partial ^i\partial_i v $ can be obtained by performing
a gauge transformation from the synchronous-comoving gauge to the
Poisson gauge.  
The equivalence at first order of the Newtonian theory with the
relativistic equations for the matter density fluctuation in the
comoving gauge and the velocity in the Poisson gauge has recently been
discussed \citep{Chisari:2011iq,Green:2011wc}\footnote{The
  relativistic effect of fluids with non-zero pressure has been
  explored by \cite{Christopherson:2012kw}.} with particular emphasis
in the study of the scale-dependent bias of the large scale structure  
(LSS) \citep{Wands:2009ex,Baldauf:2011bh,Bru12}. In the present work we
have provided 
equations that extend the correspondence to non-linear order and
establish a framework in which the effects of primordial non-Gaussianity
in the LSS can be studied, keeping the consistency within general
relativity. 

We have presented solutions up to second order in a perturbative
expansion about a background $\Lambda$CDM cosmology. %
We believe that our derivation using the fluid-flow approach is more
transparent than earlier derivations, revealing the essential role of
the GR constraint equations, relating the initial density to the
primordial curvature perturbation. 
We find what we believe to be a new, non-separable, second-order
solution for the growing mode of the local density perturbation at a
general point in the density field, with arbitrary shape coefficient,
$\Sigma$ defined in \eref{sigma:def}. 
This reduces to previously known separable solutions in the
matter-dominated (Einstein-de Sitter) limit \citep{Bar05} or in the
special case of planar symmetry ($\Sigma=1$) \citep{Bar10}. 
It would therefore be interesting to connect our results to
alternative approaches to study non-linear effects in the density
field, in particular, attempts to consider the Zel'dovich approximation in GR
\citep{Russ:1995eu,MatTer96,Villa:2011vt,Rampf:2012pu} and second-order
Lagrangian perturbation theory (2LPT) \citep{Bernardeau:2001qr}.
The essential difference between Newtonian theory and GR is
action-at-a-distance vs. causality, i.e. in the way non-locality comes
in. The gradient of \eref{Euler} would lead to the evolution equation
for $\vartheta^i_{\rm{N} j}$, \eref{Ch6:eq:evol2}, with the curvature
terms replaced by the tidal field
\citep{barrow:89,maartens-bruni}. The Zel'dovich approximation results
in truncating these equations, thereby reintroducing locality in the
evolution system. It would be interesting to explore this method of
solution of the non-linear equations, which goes beyond our
second-order perturbative expansion, in the GR context. 

The general solution at second order includes the Newtonian non-linear
growing mode $D_{2+}$ ($D_{2+} \propto D_+^2$ in the matter-dominated era),
which is known to generate a non-zero galaxy bispectrum
\citep{Jeong:2009vd}. The equivalence of the relativistic
and Newtonian non-linear evolution equations has been reported before
\citep{MatTer96,Noh:2004bc}, but there is also a second-order correction to
the linearly growing mode, $\propto D_+$, due to the non-linear
constraint equations in GR \citep{Bar05,Bar10}. This needs to be
included in the initial conditions used for N-body simulations which
then use Newtonian equations of motion to follow the evolution of
structure.   

This intrinsic non-Gaussianity in GR leads to a galaxy bispectrum
equivalent to a primordial non-Gaussianity parameter $f_{\rm NL}=-5/3$
in the squeezed limit \citep{Verde:2009hy}. 
This result is obtained by using the primordial curvature
perturbation, $\zeta$, to set initial conditions on scales larger than
the horizon length at the start of the matter era. On smaller scales
we would need to include the effect on the comoving curvature
perturbation of evolution during the preceding radiation era. This is
a more challenging calculation, especially if one considers the effect
of photon pressure on the baryons before decoupling, almost certainly
requiring a numerical calculation. Fortunately numerical codes have
recently been constructed to calculate the intrinsic non-Gaussianity
in the CMB anisotropies from second-order effects
\citep{Pitrou:2010sn,Huang:2012ub,Su:2012gt,pettinari:13} and these
codes should also be able to calculate the second-order density
perturbation at the start of the matter era, and hence the expected
galaxy bispectrum from second-order terms on intermediate scales. 

To extend our GR calculations to higher order in a perturbative
expansion would require us to include vorticity and gravitational
waves. Although vector and tensor modes will
inevitably be generated from first-order scalar perturbations
\citep{MatMolBru98,Ananda:2006af,Lu:2007cj}, they do not affect the
(scalar) density perturbation at second order. Nonetheless they do
appear in the second-order metric and need to be included in a
consistent, relativistic treatment of observable effects, such as
frame-dragging \citep{Bruni:2013mua}.

\section{Acknowledgements}
\label{ack} 
The authors are grateful to Rob Crittenden, Roy Maartens, Marc Manera,
Sabino Matarrese, Francesco Pace and Cornelius Rampf for useful
comments. JCH is funded by CONACyT (programme  \emph{Estancias
  Posdoctorales y Sab\'aticas al Extranjero para la Consolidaci\'on de
  Grupos de Investigaci\'on}). This work was supported by STFC grants
ST/H002774/1 and ST/K00090X/1. 

{Note added in proof}: {We also thank Claes Uggla and John
  Wainwright, who contacted us after this paper was submitted,
  providing useful comments. Their paper \citep{Uggla:2013kya}
  presents an interesting alternative derivation of the second order
  relativistic perturbation \eref{delta2final} in a gauge-invariant fashion.}

\appendix
\section{Solution in Eulerian coordinates}
\label{app:B}

In this appendix we make explicit the correspondence between the
second-order GR solutions \eref{delta2final} and the well-known
Newtonian version in Eulerian coordinates. In Sec. \ref{Ch6:sec:NGR}
we have shown how the equations relevant to our analysis match those of the
Newtonian treatment when the following correspondences are assumed,  
\beq
 \vartheta^i_j \to \partial^i\partial_j v \quad \mathrm{and}
\quad \frac{\partial}{\partial \tau} \to \frac{d}{d \tau} =
\frac{\partial}{\partial \tau} + \partial^i v \partial_i
\,.     
\label{newtonian:theta}
\eeq

\noindent The time derivative transformation in the Newtonian limit
from a Lagrangian or convective derivative, to the partial or Eulerian
derivative, represents a change of the spatial coordinates. In the
passive approach of cosmological perturbation theory
\citep[e.g.][]{MalWan09}, this is a change of spatial coordinate (or
threading) for a fluid element 
\beq
 \label{spatialgauge}
\bx \to \widetilde{\bx} = \bx - {\bf \xi}
\,, 
\eeq
leading to a first-order change in the 3-velocity
\beq
{\bf v}^{(1)} \to \widetilde{{\bf v}}^{(1)}= {\bf v}^{(1)} - {\bf \xi}'  \,.
\eeq

\noindent To transform from the synchronous-comoving gauge, where
${\bf v}^{(1)}=0$, to a new 
\emph{Eulerian}  gauge where $\widetilde{\bf v}^{(1)}=\nabla v_E$ we have 
\beq
\xi_i = - \int \partial_i v_{E} \,d\tau.
\eeq
In particular we choose $\xi$ such that $\nabla^2 v_E=\vartheta$.

This gauge transformation does not affect the first-order density
contrast, since scalar perturbations are invariant under first-order
spatial gauge transformations: $\delta_E^{(1)} = \delta_{\rm
  sync}^{(1)}$.  
However, at second-order, under a first-order spatial gauge
transformation \eref{spatialgauge}, we have
\citep{MatMolBru98,MalWan09} 
\beq
\delta^{(2)} \to \widetilde\delta^{(2)} = \delta^{(2)} + 2\xi^i\partial_i\delta^{(1)}
 \,,
\eeq
hence the second-order density contrast in the Eulerian gauge is
\beq
\delta_E^{(2)} = \delta^{(2)} - 2 \partial_i\delta^{(1)}  \int
\partial^i v_{E}\,d\tau 
\,.
\eeq

In terms of the first-order solutions Eqs.~(\ref{delta1final})
and~(\ref{theta1:Rc1}), which gives 
\beq
\int v_E d\tau = \int \nabla^{-2} \vartheta d\tau
 = - \left[ f_1(\Omega_m)+\frac32 \Omega_m \right]^{-1} \frac{\Rc}{\Hs^2}
\,,
\eeq
we find the Eulerian solution
\beq
\label{delta:Euler}
\delta_E^{(2)} = \delta^{(2)} + 2 \left[ f_1(\Omega_m)+\frac32
  \Omega_m \right]^{-2} \frac{\partial_i\Rc \partial^i
  \nabla^2\Rc}{\Hs^4}  
\,.
\eeq
It is clear that non-linear terms resulting from the gauge
transformation from Lagrangian to Eulerian density, will not modify
the relativistic correction, proportional to $D_+$ in
\eref{soln:delta2h} at second order, but rather contributes an
additional ``Newtonian'' term proportional to $D_+^2$. 

We can compare this with the usual Newtonian solution in the
matter-dominated limit ($\Omega_m=1$) where the full Eulerian
solution, Eqs.~(\ref{delta2final}) with (\ref{delta:Euler}), reduces
to 
\beqa  
 \label{delta2EdS}
\delta^{(2)}_E(\bx,\tau) &=&
 -\frac{24}{25\Hs^2} 
\left\{  \left( f_{\rm NL} - \frac53 \right) \Rc\nabla^2\Rc + \left( f_{\rm NL} + \frac{5}{12} \right) \partial^j\Rc \partial_j\Rc 
 \right\}
 \nonumber\\ 
&& +
\frac{8}{25\Hs^4}
\left\{
\frac57 \left( \nabla^2\Rc \right)^2 + \frac27 \partial^i\partial_j\Rc \partial_i\partial^j\Rc + \partial_i\Rc \partial^i \nabla^2\Rc
\right\}
\,.
\eeqa

\noindent Transforming to Fourier space we obtain
\beq
\delta^{(2)}_{E{\bf k}} = 2 \int \frac{d^3{\bk}_1\, d^3{\bk}_2}{(2\pi)^3} \delta_D({\bk}-{\bk}_1-{\bk}_2)
 \, F_2({\bk}_1,{\bk}_2) \delta^{(1)}_{{\bk}_1} \, \delta^{(1)}_{{\bk}_2} \,,
\eeq

\noindent with the kernel
\beqa
\label{F2}
 F_2({\bk}_1,{\bk}_2) &=&
  - \frac{3}{D_+} \left\{ \left( f_{\rm NL} - \frac53 \right) 
  \frac{k_1^2+k_2^2}{2k_1^2k_2^2} 
+ \left( f_{\rm NL} + \frac{5}{12} \right) 
\frac{{\bk}_1\cdot{\bk}_2}{k_1^2k_2^2} \right\}
\nonumber
\\
&&   + \left\{ \frac57 + \frac27 
\frac{\left({\bk}_1\cdot{\bk}_2\right)^2}{k_1^2k_2^2} 
+ \frac{{\bk}_1\cdot{\bk}_2 \left(k_1^2+k_2^2\right)}{2k_1^2k_2^2} \right\}.
 \eeqa

\noindent The second line of this equation, which dominates at late
times, reproduces exactly the second-order Newtonian solution in Eulerian
coordinates, e.g., Eq.~(45) in section~2.4.2 of
\citep{Bernardeau:2001qr} \footnote{Note that in
  \citep{Bernardeau:2001qr} the perturbative expansion of non-linear
  variables does not carry the usual Taylor-expansion numerical
  factors. Hence the factor of $2$ included in our definition of $F_2$
  in \eref{F2}.}.

The first line of this equation represents the non-linear initial conditions in GR, including both primordial non-Gaussianity and intrinsic non-Gaussianity due to non-linear constraints in GR.
In the squeezed limit, $k_1\to0$, the first term dominates and we have
\beq
F_2 \to - \frac{3}{2D_+} \left( f_{\rm NL} - \frac53 \right) \frac{1}{k_1^2}
 \,,
\eeq
showing the effect of GR corrections as an effective shift in the
value of $f_{\rm NL}$, $\Delta f_{\rm NL}=-5/3$.



\end{document}